\documentclass[12pt]{article}
\usepackage{epsfig}
\usepackage{amstex}

\newlength{\hepwidth}
\newcommand{\psibar} {\ensuremath{\bar{\psi}}}

\newcommand{\scr}[1] {\mbox{\scriptsize #1}}
\newcommand{\mq} {\ensuremath{m_{\scr{q}}}}

\newcommand{\tc} {\ensuremath{T_{\scr{c}}}}

\def\PR{{\sl Phys.\ Rev.\ }} 
\def\NP{{\sl Nucl.\ Phys.\ }} 

\begin{document}
\hfill\parbox{\hepwidth}{
IFUP-TH 40/96\\
BI-TP 96/27\\
hep-lat/9607046}

\vspace{1cm}
\begin{center}
  {\Large\bf Two Flavour QCD Phase Transition\\[1ex]}
  G.~Boyd$^a$\footnote{Talk presented at the 10th
    International Conference {\em Problems of Quantum
      Field Theory } Alushta (Crimea, Ukraine)  13 - 18 May 1996},\\
  with F. Karsch$^b$, E. Laermann$^b$ and M. Oevers$^b$ \\[1ex]
  (a) Dipartimento di Fisica dell'Universit\`a,
  I-56126 Pisa, Italy \\
  (b) Fak.\ f\"ur Physik, Uni.\ Bielefeld, Postfach 100131, D-33501
  Bielefeld, Germany \\
\end{center}

\begin{abstract}
  Results on the phase transition in QCD with two flavours of light staggered
  fermions from an ongoing simulation are presented.  We find the restoration
  of the chiral $SU(2)\times SU(2)$ symmetry, but not of the axial
  $U_{\!\text{A}\!}(1)$ symmetry.
\end{abstract}


\section{Deconfinement and Chiral Symmetry}
There are two symmetries in QCD, each giving rise to a phase transition. For
infinite quark mass, ie., pure $SU(3)$ gauge theory, there is the deconfinement
transition.  The order parameter is the Polyakov loop $\langle L\rangle$, zero
in the confined Z(3) symmetric phase, non-zero in the deconfined broken
symmetry phase.  For zero quark mass one has chiral symmetry, the chiral
condensate $\langle\psibar\psi\rangle$ the order parameter.  Chiral symmetry is
spontaneously broken at $T=0$, so $\langle\psibar\psi\rangle$ is non-zero at
low temperature, and zero above the chiral transition.

In the real world with small quark masses it is not clear what role each
symmetry assumes going from cold to hot QCD.  There could be two different
transitions, with two different transition temperatures. Confinement and chiral
symmetry breaking may be related~\cite{BANKS80LEUTS92}, with one causing the
other. That they need not be coupled has been discussed in~\cite{FGM92}.  In
this project we saw deconfinement and chiral symmetry restoration at the same
temperature, with in the susceptibilities corresponding to
$\langle\psibar\psi\rangle$ and $\langle L\rangle$ peaking at the same
temperature.

The order of the phase transition will depend on which symmetry dominates at
the transition. Since the quark mass is close to zero, most likely the chiral
symmetry will dominate.  For the case of two light flavours there is no clear
prediction of the order of the transition from effective
theories~\cite{RPW84RFW92}. If it turns out to be of second order, then it
should belong to the same universality class as the O(4) spin model.

The lattice calculations of the order of the transition presented here
extend earlier work~\cite{KRHSUS94LAER94} (where details of the
method may be found) to larger lattices. See also~\cite{UKAWA96}.

The bare quark mass in the QCD Langrangean behaves analogously to a magnetic
field in a spin system, imposing an upper limit on correlation lengths.  Of
course, the physical size of the system must be larger than this limit.  The
order and universality class of the phase transition can then be determined by
calculating the behaviour of various correlation lengths and corresponding
susceptibilities as the quark mass is taken to zero near the deconfinement
transition. For a second order chiral phase transition one expects the chiral
susceptibility $\chi_{m}$ to grow without bound when the quark mass is taken to
zero at the critical temperature, while for a first order transition it will
grow to some maximum value.

\begin{equation}\begin{aligned}
\chi_{m}(\mq) & = \frac{T}{V}\frac{\partial}{\partial\mq}
                                            \langle\psibar\psi\rangle \\
              & = \frac{T}{V}\frac{\partial^{2}}{\partial\mq^{2}}
                                            \ln(Z)
\end{aligned}\end{equation}
$Z$ is the QCD partition function. Moving across the phase transition
at a given quark mass \mq, the chiral susceptibility rises to a maximum,
and then drops. The height of the peak depends on the quark mass via the
critical exponent $\delta$
\begin{equation}
\chi_{m,\text{peak}} = c_{m}\mq^{1/\delta - 1}.
\label{eq:susm}
\end{equation}

\begin{figure}[t]
  \begin{center}
    \leavevmode
    \epsfig{file=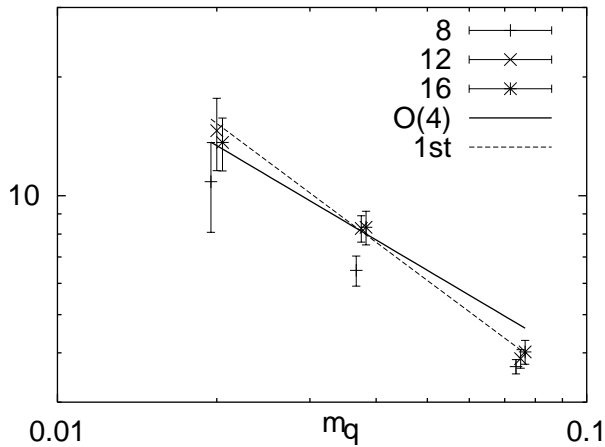}
    \caption{The chiral susceptibility $\chi_{m,\text{peak}}$  at the 
      deconfinement transition vs. \mq for three different lattice volumes,
      $N_{\sigma}^{3}\times 4$. The points have been shifted slightly for ease
      of comparison. Also shown is the slope expected for a second order
      transition with O(4) exponents, and the slope expected for a first order
      transition.}
    \label{fig:susm}
  \end{center}
\end{figure}
In figure~\ref{fig:susm} the peak height is plotted against quark mass for
three different lattice sizes, $8^{3}, 12^{3}$ and $16^{3}\times 4$ and for
three different quark masses, $\mq=0.02, 0.0375$ and $0.075$.  Also plotted are
two lines. One represents the slope expected if one has a second order
transition with O(4) exponents, $1/\delta=0.206$~\cite{KANKAY}, the other if
there is a first order transition with $1/\delta=0$.  Both exponents fit the
data more or less well, with larger lattices suggesting more that $1/\delta=0$.
However, the scaling arguments behind eq.~\eqref{eq:susm} may well not hold for
the heavy mass $\mq=0.075$. If this mass is ignored then no conclusion can be
drawn.



\section{Axial Symmetry}
There has been speculation about the possible restoration of the
$U_{\!\text{A}\!}(1)$ axial symmetry at high temperature~\cite{SHURUA1}.
Although the anomaly explicitly breaks the symmetry regardless of temperature,
it is only present in one loop diagrams, and thus the effect may disappear as
the temperature is increased.  Examining the finite temperature masses of
mesons with a mass splitting due to the axial anomaly, this effective
restoration of the axial symmetry may be studied.

\begin{figure}[t]
  \begin{center}
    \leavevmode
    \epsfig{file=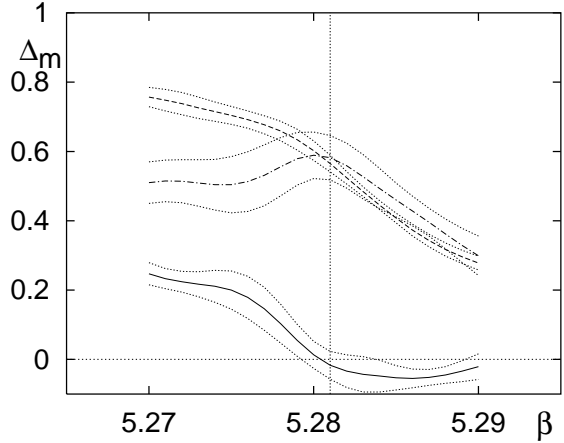,height=6cm}
    \caption{The mass differences $m_{\sigma}-m_{\pi}$ (solid line),
    $m_{\delta}-m_{\pi}$ (dashed line) and $m_{\delta}-m_{\sigma}$ (dash-dotted
    line). The vertical line represents the critical $\beta$, the range of
    $\beta$ values spans a few \% on either side of \tc. }
    \label{fig:axial}
  \end{center}
\end{figure}

The particles to consider are the pion ($\pi$), the scalar isoscalar ($\sigma$)
and the neutral component of the scalar isovector ($\delta$), the $\delta^{0}$.
The $\sigma$ and $\delta^{0}$ differ only in diagrams with disconnected quark
lines. In other words, the quark line connected diagrams are identical for both
particles, and the disconnected diagrams give the $\sigma$ a new set of poles.

If chiral $SU(N)\times SU(N)$ symmetry is restored, the pion and sigma will be
degenerate. If both  $SU(N)\times SU(N)$ and $U_{\!\text{A}\!}(1)$ is restored,
then all three will be degenerate. 
The following mass splittings
\begin{equation}\begin{gathered}
    \Delta_{\pi\sigma} = m_{\sigma} - m_{\pi} \\
    \Delta_{\pi \delta} = m_{\delta} - m_{\pi} \\
    \Delta_{\delta\sigma} = m_{\delta} - m_{\sigma}
\end{gathered}\end{equation}
can be used to determine which symmetry is restored, and at which
temperature. 

The short temporal direction at finite temperature prevents an easy
determination of particle masses from a fit of an exponential to the
correlator. However, one can also link the susceptibility, or correlator at
zero four-momentum, to the mass:
\begin{equation}
1/m^{2} \propto \chi = \int d^{4}x C(x).
\end{equation}
This definition assumes that only one state contributes to the
correlator. If the mass gap is large, and the lowest state has a mass
close to zero, then the bias introduced by using the full correlator will not
be large. 

This definition of the mass has been used in calculating the differences shown
in figure~\ref{fig:axial}. The figure shows quite clearly that although
$\Delta_{\pi\sigma}$ goes to zero at \tc, neither $\Delta_{\pi \delta}$ nor
$\Delta_{\delta\sigma}$ do\footnote{For technical reasons the $\delta$ only
  corresponds to the physical $\delta$ if the lattice is close to the continuum
  limit.}. So chiral symmetry has been restored, while $U_{\!\text{A}\!}(1)$
has not. Both $\Delta_{\pi \delta}$ and $\Delta_{\delta\sigma}$ do drop slowly
to zero as the temperature increases, though, indicating an effective
restoration of $U_{\!\text{A}\!}(1)$. See also~\cite{DETAR96}.

So one concludes that at \tc\ chiral $SU(N)\times SU(N)$ symmetry is restored
and deconfinement takes place, while $U_{\!\text{A}\!}(1)$ symmetry is not
restored. At some temperature above \tc, $U_{\!\text{A}\!}(1)$ becomes
effectively restored. Finally, note that the order of the phase transition is
still undecided.

\section*{Acknowledgements}
GB acknowledges partial support from the European Union Human Capital and
Mobility program HCM-Fellowship contract ERBCHBGCT940665. The work has been
made possible by the 256 node QUADRICS parallel computer funded by the DFG
under contract no. Pe 340/6-1 for the DFG-Forschungsschwerpunkt "Dynamische
Fermionen", as well as the generous provision of time on the 512 node QUADRICS
computer of ENEA.



\end{document}